\documentclass[runningheads]{llncs}
\usepackage[T1]{fontenc}
\usepackage{graphicx}

\usepackage{cite}
\usepackage{amsmath,amssymb,amsfonts}
\usepackage{algorithmic}
\usepackage{graphicx}
\usepackage{textcomp}
\usepackage{makecell}
\usepackage{booktabs}
\usepackage{enumitem}
\usepackage{float}

\def\mathbi#1{\textbf{\em #1}}

\def\mathbi#1{\textbf{\em #1}}

\begin{document}
\title{Frequency-aware Convolution for Sound Event Detection}
%
%

\author{Tao Song\inst{1}$^{\star}$ Wenwen Zhang\inst{2}\thanks{These authors contributed to the work equally and should be regarded as co-first authors.}}

%
\authorrunning{Tao Song, Wenwen Zhang}
%
\institute{Kuaishou Technology, Beijing, China \and Beijing University of Posts and Telecommunications, Beijing, China}
%
\maketitle              
%
\begin{abstract}
In sound event detection (SED), convolutional neural networks (CNNs) are widely employed to extract time-frequency (TF) patterns from spectrograms. However, the ability of CNNs to recognize different sound events is limited by their insensitivity to shifts of TF patterns along the frequency dimension, caused by translation equivariance. To address this issue, a model called frequency dynamic convolution (FDY) has been proposed, which involves applying specific convolution kernels to different frequency components. However, FDY requires a significantly larger number of parameters and computational resources compared to a standard CNN.
This paper proposes a more efficient solution called frequency-aware convolution (FAC). FAC incorporates frequency positional information by encoding it in a vector, which is then explicitly added to the input spectrogram. To ensure that the amplitude of the encoding vector matches that of the input spectrogram, the encoding vector is adaptively and channel-dependently scaled using self-attention.
To evaluate the effectiveness of FAC, we conducted experiments within the context of the DCASE 2023 task 4. The results show that FAC achieves comparable performance to FDY while requiring only an additional 515 parameters, whereas FDY necessitates an additional 8.02 million parameters. Furthermore, an ablation study confirms that the adaptive and channel-dependent scaling of the encoding vector is critical to the performance of FAC.

\end{abstract}
\section{Introduction}
Sound event detection (SED) aims to recognize what is happening in an audio clip and when it is happening. As a significant research topic in computational auditory scene analysis, SED has garnered considerable attention in recent years\cite{Hyeonuk_2022, Ebbers_2022_Threshold, Li_2023_AST, Xu_2023_Semi, Shao_2024_Fine, Li_2024_Semi}. There is an international challenge named DCASE, which focuses on SED and has been held every year since 2016\footnote{\url{https://dcase.community}}.

SED can be formulated as a \textit{multi-class, multi-label classification} problem. In a typical deep learning-based system, the input audio clip is first converted into a spectrogram, which is then used to extract time-frequency (TF) patterns. These patterns are subsequently classified by a deep neural network (DNN) to detect sound events within each frame. By combining consecutive frames with the same event class, the onset and offset of each sound event can be determined.

The frequency position of TF patterns within a spectrogram plays a crucial role in SED. Under certain circumstances, shifting a TF pattern along the frequency axis can result in a perceived difference or even misclassification as another sound event. For instance, the sound events \textit{Blender} and \textit{Vacuum cleaner} are primarily produced by operating motors and share similar TF patterns. However, due to the higher rotation speed of a blender compared to a vacuum cleaner, the TF patterns of the Blender are positioned at higher frequencies. Shifting the TF patterns of the \textit{Blender} downwards along the frequency axis may cause them to be perceived as \textit{Vacuum cleaner}.

%
\begin{figure}[b!]
	 \centering
	 \includegraphics[width=\linewidth]{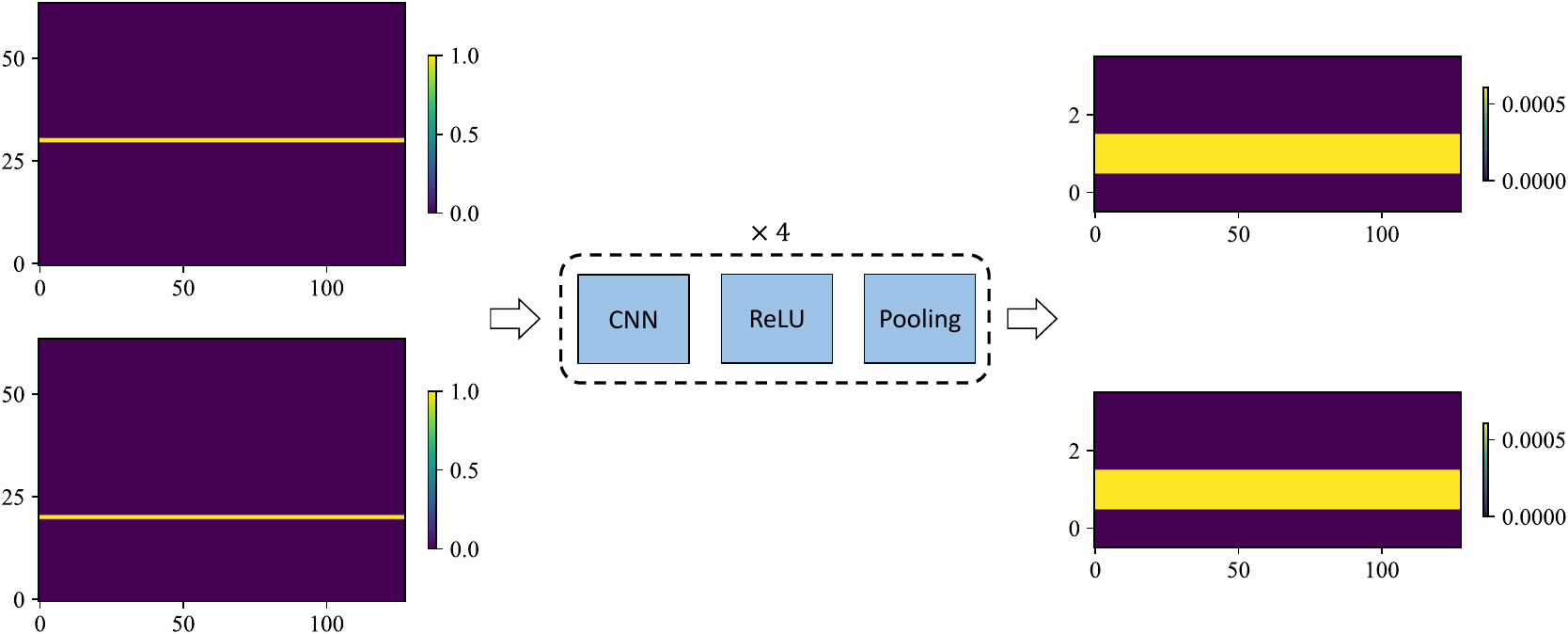}
	\caption{Example of the insensitivity of a CNN module to shift of time-frequency patterns. The input spectrogram has 64 frequency bins, with only one bin has a nonzero value of 1. Shifting the TF pattern by 10 frequency bins does not change the output of the CNN module.}
	\label{fig:eg_CNN_shift_insensitive}
\end{figure}
In SED, a convolutional neural network (CNN) module is typically employed to extract TF patterns from the spectrogram\cite{Cakir_2017, Miyazaki_2020, Ebbers_2021, Guirguis_2021, Wakayama_2022}. The CNN module consists of multiple CNN layers, each CNN layer is followed by a pooling layer to reduce the frequency dimension of the spectrogram. However, due to the translation equivariance of CNNs and the translation invariance of pooling layers, the features extracted may lack sensitivity to shifts of TF patterns along the frequency axis. An illustrative example is given in the Fig. \ref{fig:eg_CNN_shift_insensitive}. The input spectrogram has $64$ frequency bins, with only the $20th$ bin having a non-zero value of 1. A CNN module is constructed by concatenating $4$ CNN blocks, where each block consists of a CNN layer followed by a pooling layer. All CNN kernels share a shape of $3\times3$, and ReLU activation is applied to the output of each CNN layer.  In the pooling layer, average pooling is applied only in the frequency dimension, with a window size of 2. If we shift the TF pattern upward by 10 frequency bins, the output of the DNN model remains unchanged\footnote{The maximum shift that the DNN model can tolerate depends on many factors, such as the kernel value and the activation function.}. The CNN module has a limited sensitivity to shifts in TF patterns along the frequency dimension. In other words, the CNN module is not sensitive to the frequency position of the TF pattern.

In SED, the performance of a CNN-based model may be limited by its insensitivity to the frequency position of TF patterns. To address this issue, a frequency dynamic convolution (FDY) approach has been proposed in \cite{Hyeonuk_2022}. FDY applies specified convolution kernels to different frequency components of the input spectrogram, and as a result, features extracted would vary as TF patterns shift along the frequency dimension. Although FDY achieves notable performance improvements, it requires several times more trainable weights and calculations compared to a vanilla CNN. 
This paper introduces a more efficient solution called frequency-aware convolution (FAC). FAC encodes frequency-positional information into a vector and adds it to the input spectrogram. To ensure compatibility with the amplitude of the input spectrogram, the encoding vector is adaptively and channel-dependently scaled using self-attention.

The rest of this paper is organized as follows: Section 2 discusses related works, Section 3 describes the proposed method, Section 4 presents the experimental setup and evaluation results, and Section 5 concludes the paper.

\section{Related work}

Frequency positional information, in addition to SED, is also important in other audio processing tasks, such as acoustic scene classification. Several studies have explored the extraction of frequency position-aware features using CNNs\cite{Khaled_2019, Rakowski_2019, Hyeonuk_2022, Nisan_2023}. 


In \cite{Khaled_2019}, a Frequency-aware CNN is proposed, where a feature map encoding the frequency positional information is concatenated to the input spectrogram along the channel dimension. The feature map is defined as 
\begin{equation}
\mathbi{V}(t, f) = f/F
\end{equation}
where $t$ is the time, $f$ is the frequency, and F is the size of the frequency dimension of the feature map. The output of CNN can be formulated as 

\begin{equation}
\begin{aligned}
\label{eq:FACNN}
\mathbi{Y'} &= \text{Conv2d}([\mathbi{X}, \mathbi{V}], \mathbi{k})  \\
            &= \text{Conv2d}(\mathbi{X}, \mathbi{k}_1) + Conv2d(\mathbi{V}, \mathbi{k}_2) \\
            &= \mathbi{Y} + \mathbi{e}
\end{aligned}
\end{equation}
where $\mathbi{X}$ is the input spectrogram, $\mathbi{k}$ is the convolution kernel. $\mathbi{Y}$ is the original output of CNN, and $\mathbi{e}$ can be regarded as an encoding of frequency positional information. As a result, this method is equivalent to directly adding a trainable vector to the original CNN output. However, there is a problem that $\mathbi{Y}$ and $\mathbi{e}$ may mismatch in amplitude, given that the amplitude of the spectrogram can be very fluctuant.

The insensitivity of a CNN module to the frequency positional information results from the combination of CNN and pooling layers. \cite{Rakowski_2019, Nisan_2023} proposed a simple solution where no pooling operation is applied to the frequency dimension. However, this may result in extracted features that are too large to be processed by following modules, despite the increased calculation.

The frequency dynamic convolution (FDY) is proposed in \cite{Hyeonuk_2022}, which extracts frequency-aware features by applying specified kernels to different frequency components. FAC uses $K$ basis kernels and frequency-adaptive attention weights to generate frequency-adaptive kernels, 
\begin{equation}
\begin{aligned}
\mathbi{W} &= \sum_{i=1}^{K}{\mathbf{\pi}_i(f, \mathbi{X}) \times \mathbi{W}_i} \\
b &= \sum_{i=1}^{K}{\mathbf{\pi}_i(f, \mathbi{X}) \times b_i} \\
\end{aligned}
\end{equation}
where $\mathbf{W}_i$ and $b_i$ are weight and bias for $i$th basis kernel, $\mathbf{\pi}_i(f, \mathbi{X})$ is frequency-adaptive attention weight for $i$th basis kernel. The final output of FAC can be calculated as  
\begin{equation}
\begin{aligned}
\mathbi{Y}(t, f) &= \mathbi{W}*X(t, f) + b \\
                 &= \sum_{i=1}^{K}{\mathbf{\pi}_i(f, \mathbi{X}) \times \left(\mathbi{W}_i * \mathbf{X}(t, f)+b_i\right)}\\
                 &= \sum_{i=1}^{K}{\mathbf{\pi}_i(f, \mathbi{X}) \times \mathbi{Y}_i}
\end{aligned}
\end{equation}
FAC requires $K-1$ times more convolution kernels and convolution operations than a vanilla CNN, which greatly increases the amount of trainable weights and calculations. 

In summary, the frequency position of TF patterns is an important feature for detecting sound events. However, CNN modules are insensitive to this information, and existing solutions often significantly increase the model's parameter count and computational complexity.
\section{Proposed method}
\label{sec:method}

To enhance the sensitivity of CNN modules to frequency positional information, a straightforward approach is to explicitly encode this information into the input. Assume the input spectrogram $X$ has $C$ channels, $T$ frames, and $F$ frequency bins. A vector $P_{freq}$ of length $F$ can be used to encode frequency positional information. 
\begin{equation}
\mathbi{P}_{freq}(f) = \sin(\frac{\pi}{2} \times f/F)
\end{equation}
To match the input size, $\mathbi{P}_{freq}$ is replicated $T$ times along the time axis and $C$ times along the channel axis. In other words, the same encoding vector is shared by each frame and each channel. This vector is added to each frame of $X$, resulting in the modified input $X'$ for the CNN layer. 
\begin{equation}
	\mathbi{X}'(t, f, c) = \mathbi{X}(t,f,c) + \mathbi{P}_{freq}(f) 
\end{equation}

where   $t$ is the time, $f$ is the frequency, and $c$ is the channel. The encoding vector $P_{freq}$ can be learned by the model. If $X$ is shifted along the frequency dimension by $\Delta f$, $X'$ will change as long as $P_{freq}(f) \neq P_{freq}(f + \Delta f)$, thus making the CNN module sensitive to frequency positional information.

However, this method has limitations. The amplitude of the spectrogram varies over time and across channels. For different frames and channels of $X$, the amplitude of the frequency positional encoding vector might be too large or too small. If the amplitude is too large, it may mask the spectral information. Conversely, if it is too small, it might not effectively convey the frequency positional information. 
\begin{figure}[htbp]
	 \centering
	 \includegraphics[width=\linewidth]{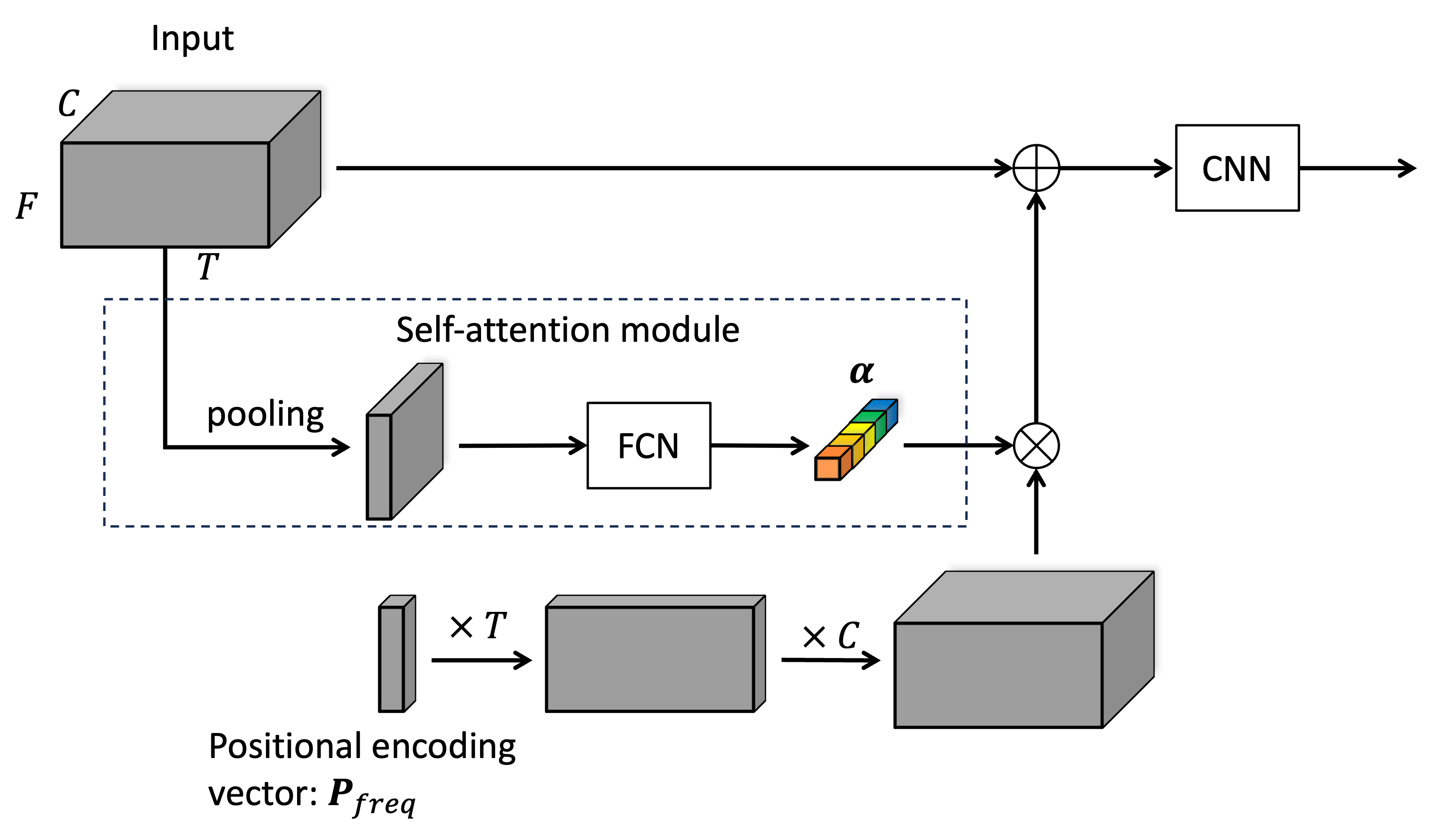}
	\caption{Frequency-aware convolution.}
	\label{fig:freq_aware_CNN}
\end{figure}

To address this issue, this paper proposes the frequency-aware convolution (FAC), which dynamically balances the amplitude of the frequency positional encoding and the input spectrogram using self-attention. Fig. \ref{fig:freq_aware_CNN} illustrates the basic structure of FAC. In FAC, the input $X'$ to the CNN layer is formulated as, 
\begin{equation}
\begin{aligned}
\mathbi{X}'(t, f, c) &= \mathbi{X}(t,f,c) + \mathbf{\alpha}(c) \times \mathbi{P}_{freq}(f) \\
\mathbf{\alpha}(c) &= \text{NN}_{self\_attention}\left(\mathbi{X}(t,f,c)\right)
\end{aligned}
\end{equation}
where $\alpha \in [0, 1]$ is the scaling coefficient of the frequency positional encoding vector, independently determined for each channel, and $\text{NN}_{self\_attention}$ is the self-attention module. $\alpha$ is estimated from the input spectrogram: first, the time dimension of the input is compressed using average pooling; then, a single-layer fully connected network (FCN) estimates the amplitude coefficients for each channel. The FCN has $C$ neurons with a sigmoid activation function. When $\alpha = 0$, the frequency positional encoding has no effect, and FAC degenerates into a traditional convolution. When $\alpha = 1$, the attention mechanism has no effect, and the frequency positional encoding is directly added to the input.

In summary, this paper introduces the FAC with two main innovations: 1) encoding the frequency positional information and adding it into the input, and 2) using self-attention to dynamically balance the relative amplitude between the positional encoding vector and the input spectrogram. Compared to a traditional CNN layer, FAC introduces only one learnable encoding vector of length $F$ and a single FCN layer, resulting in reduced parameter count and computational complexity.

\section{Experiment}
Two experiments are carried out in the context of DCASE 2023 task4. The first experiment compares the performance of the proposed model with FDY. The second experiment explores the importance of adaptively and channel-dependently scaling the encoding vector.


\label{sec:experiment}
\subsection{Experimental setup}


In \cite{Hyeonuk_2022}, FDY is proposed and applied to a convolutional recurrent neural network (CRNN). This CRNN is modified from the baseline\footnote{\url{https://github.com/DCASE-REPO/DESED\_task/tree/master/recipes/\\dcase2023\_task4\_baseline}} of DCASE 2023 task 4 by doubling the kernel number in each CNN layer and replacing ReLU with Context Gating. The same CRNN model is used as a baseline in this work. CRNN with FDY or FAC employed is denoted as FDY-CRNN and FAC-CRNN, respectively. FAC-CRNN and FDY-CRNN are trained and evaluated under the same settings. The official implementation\footnote{\url{https://github.com/frednam93/FDY-SED}} of FDY-CRNN is adopted.

All experiments are conducted on the DESED dataset\cite{Serizel_2020}, which consists of 10 sound event classes. Models are trained using Mean Teacher\cite{Tarvainen_2017}. An early-stop strategy is adopted, which stops the training procedure if better performance is not achieved in 10 consecutive epochs. The minimum epoch number is set to 500. Log-mel spectrogram is used as the input, which is extracted on 16 kHz audio with 128 frequency bins, 2048 window length, and 256 hop length. The same data augmentation scheme from \cite{Hyeonuk_2022} is adopted, which consists of frame shift, Mixup\cite{Hongyi_2018}, time masking, and FilterAugment\cite{Nam_2022}.

Models are evaluated on the validation subset. A decision threshold of 0.5 and event-specific median filter size are used for each sound event class\cite{Hyeonuk_2022}. For each condition, an average performance of 5 runs is reported. Three performance metrics from DCASE 2023 task 4 are used, which are event-based F-score (event-F1) and polyphonic sound event detection score (PSDS)\cite{Bilen_2020} with two sets of parameters(PSDS1, PSDS2). Details for these metrics can be found in \cite{DCASE2023_task4}.

\subsection{Number of FAC layers in FAC-CRNN}

If FAC is only applied in the first block of the CNN module, the frequency positional information may gradually diminish as more blocks are added, ultimately making the output of the CNN module insensitive to frequency positional information. Therefore, it is reasonable to use FAC in as many blocks as possible. However, as the number of blocks increases, the frequency dimension of the TF representation becomes shorter, possibly diminishing the performance gains from FACs. This section primarily conducts experiments to explore the impact of the number of FAC layers on the performance of the CRNN model.

\begin{figure}[htbp]
	 \centering
	 \includegraphics[width=0.8\linewidth]{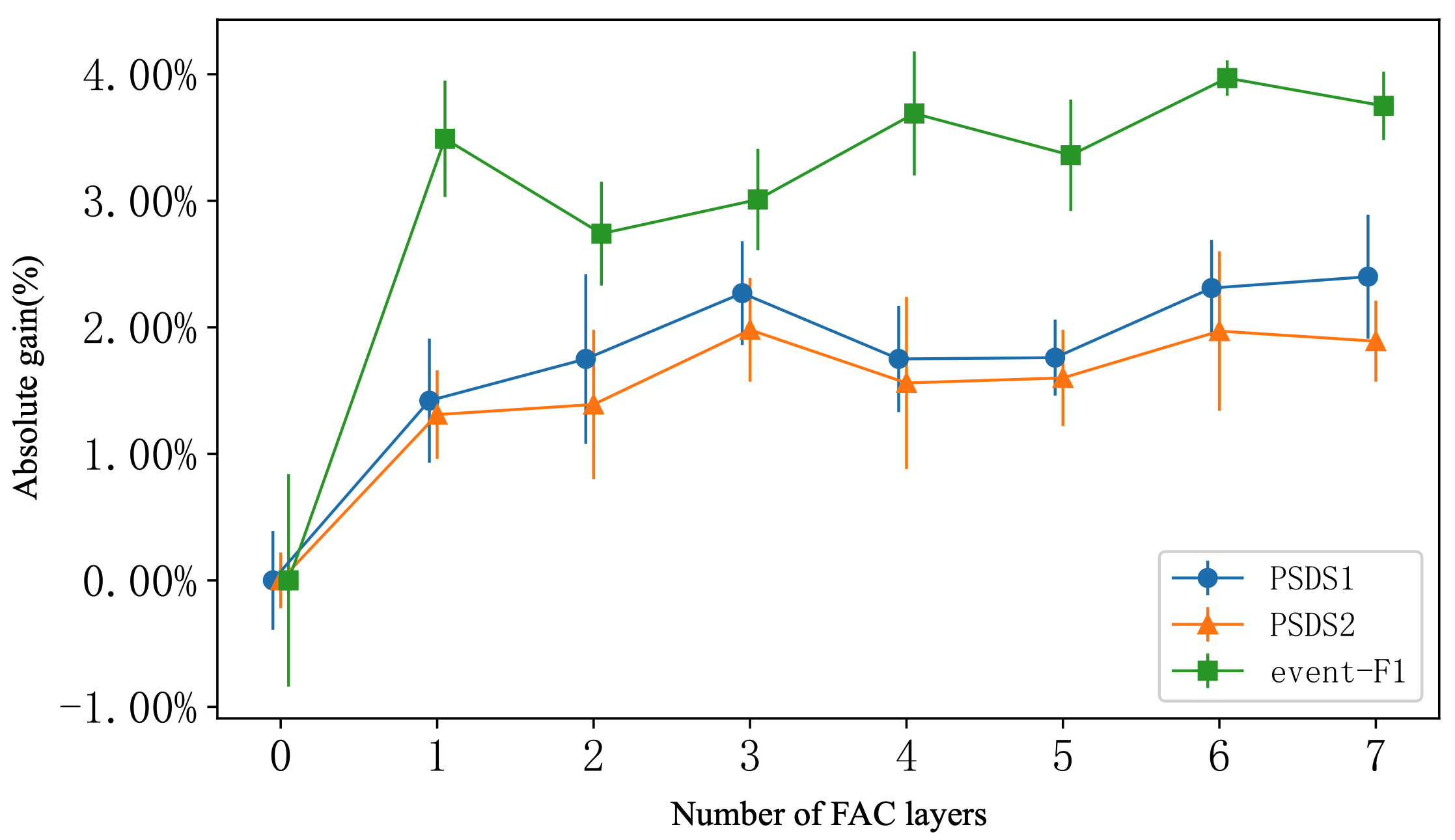}
	\caption{The curve of model performance gain with the number of FAC layers.}
	\label{fig:FAC_perform_n_layer_1}
\end{figure}

The CNN module consists of 7 convolutional blocks. Starting from the first block, the traditional CNN layers are gradually replaced with FAC layers. In this paper, and \( n_{\text{FAC}} \) denotes the number of convolutional blocks in FAC-CRNN that use FAC layers. When \( n_{\text{FAC}} = 0 \), the FAC-CRNN degenerates into a traditional CRNN model. Using CRNN model as a baseline, we calculate the performance gains of the FAC-CRNN model for various values of \( n_{\text{FAC}} \). The results are shown in Fig. \ref{fig:FAC_perform_n_layer_1}.

When \( n_{\text{FAC}} = 1 \), FAC-CRNN outperforms CRNN in all three performance metrics, demonstrating the effectiveness of FAC. As \( n_{\text{FAC}} \) increases further, the event-F1 remains relatively stable, while the PSDS1 and PSDS2 continue to improve. This indicates that as the number of convolutional blocks increases, the frequency positional information in the initial frequency convolutional block gradually diminishes, and applying FAC in subsequent convolutional blocks further improves model performance. However, the performance plateau as \( n_{\text{FAC}} \) increases. This is primarily because the frequency dimension of the TF representation has become sufficiently small. Each convolutional block in the CNN module compresses the frequency dimension by half. In the fourth block, the frequency dimension has been compressed by a factor of 8. Shifting the input of this block along the frequency dimension by one unit is equivalent to shifting the original TF spectrum by 8 frequency bands, which surpasses the variations in most sound event TF patterns. At this stage, introducing FAC no longer provides significant performance gains. In subsequent experiments in this paper, \( n_{\text{FAC}} \) of FAC-CRNN is set to 7.

\begin{figure}[htbp]
	 \centering
	 \includegraphics[width=0.75\linewidth]{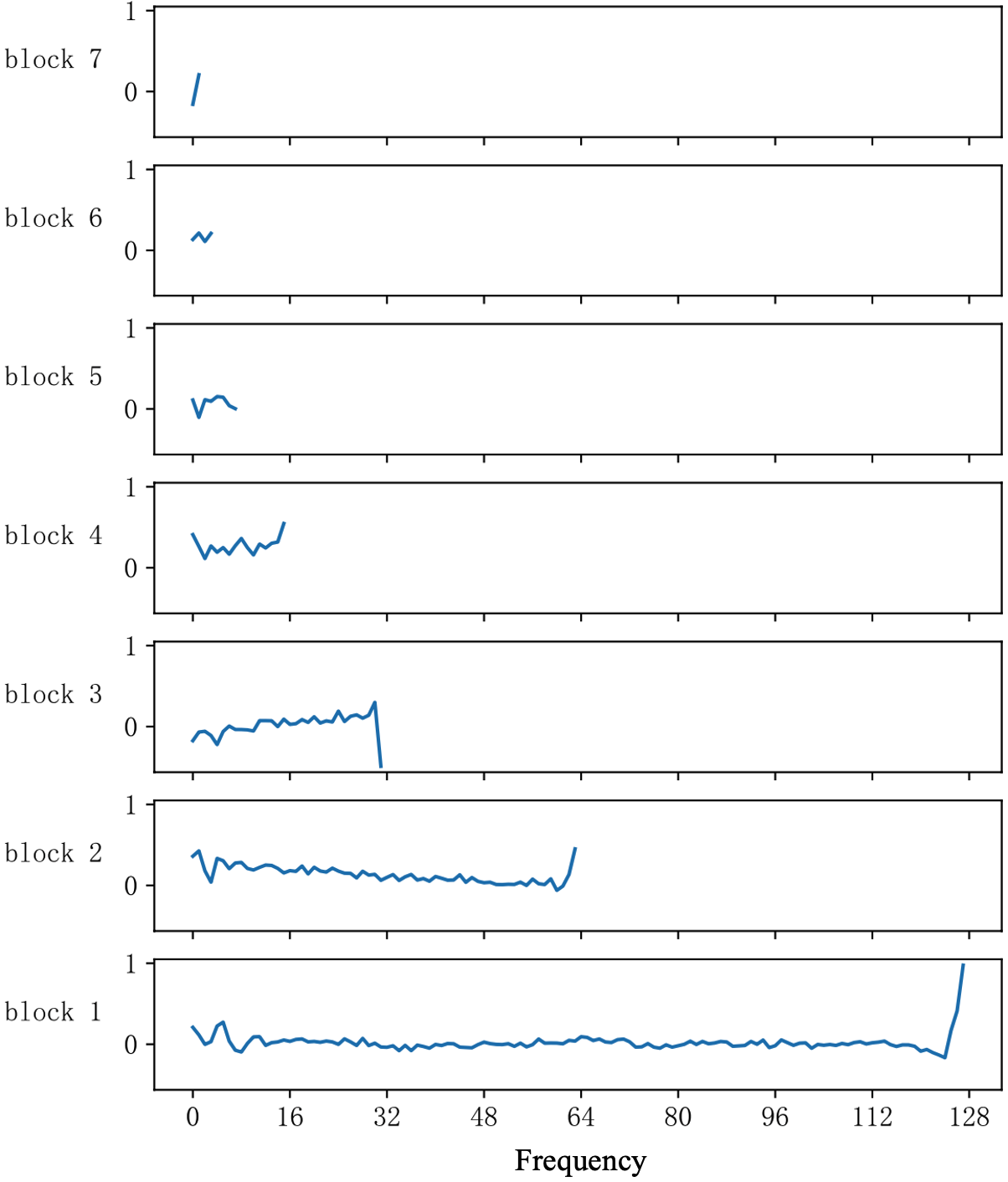}
	\caption{Encoding vector learned by each FAC layer.}
	\label{fig:FAC_perform_n_layer_2}
\end{figure}

Fig. \ref{fig:FAC_perform_n_layer_2} shows the frequency positional encoding vectors learned by 7 convolutional blocks. It can be observed that the frequency positional encodings of the first three convolutional blocks exhibit certain patterns. 
Firstly, the positional encoding vectors are monotonically increasing as a whole, enabling them to encode absolute positional information of frequency. Secondly, there is a certain periodicity in the local regions of the encoding vectors, enabling them to encode relative positional information of frequency.

\subsection{Comparasion between FAC and FDY}

The performance of CRNN, FDY-CRNN, and FAC-CRNN are listed in Table \ref{tab:result_FDY_FAC}. Firstly, both FAC-CRNN and FDY-CRNN outperform CRNN. Compared to CRNN, FAC-CRNN achieves an average relative improvement of 5.28\%, with increases of 5.61\% in PSDS1, 2.86\% in PSDS2, and 7.37\% in event-F1. These results prove the effectiveness of FAC. 
Secondly, FAC-CRNN performs similarly to FDY-CRNN in PSDS2, but slightly lower in PSDS1 and event-F1.

\renewcommand\arraystretch{1.4} 
\begin{table}[htbp]
	\centering
	\caption{Performances and model size of CRNN, FDY-CRNN, and FAC-CRNN.}
	\label{tab:result_FDY_FAC}
		\begin{tabular*}{\linewidth}{@{\extracolsep{\fill}} c|ccccc }
        \hline
    	    & \thead{PSDS1(\%)} & \thead{PSDS2(\%)} & \thead{event-F1(\%)}        & \thead{\#param(M)} & \thead{flops(G)} \\ \hline
\textbf{CRNN}     & $42.82 \pm 0.39$ & $66.31 \pm 0.22$ & $50.88 \pm 0.84$ & 3.04 & 3.35 \\
\textbf{FDY-CRNN}    & $\mathbf{46.37 \pm 0.55}$ & $\mathbf{68.70 \pm 0.51}$ & $\mathbf{55.63 \pm 0.27}$ & 11.06 & 11.64 \\
\textbf{FAC-CRNN}     & $45.23 \pm 0.49$ & $68.20 \pm 0.32$ & $54.63 \pm 0.27$ & 3.04 & 3.35 \\
 \hline
	\end{tabular*}%
\end{table}

Parameter count and flops of each model are also listed in Table \ref{tab:result_FDY_FAC}. FDY significantly increases model parameters and computation, while FAC incurs minimal additional overhead. In FDY-CRNN, frequency dynamic convolution uses 4 times the convolution kernels and convolution operations of traditional CNN, which increases model parameter count and computation by more than 2 times. FAC introduces frequency position encoding and attention on the basis of traditional CNN, which increases the complexity of the model, but these increments are negligible for CRNN.
CRNN has 3.04 million parameters and 3.35 billion flops. Compared to CRNN, FAC-CRNN only requires an additional 515 parameters and incurs an extra 1.92 million flops. It can be concluded that FAC-CRNN achieves improved performance at almost no additional cost.



\subsection{Ablation}
In FAC, a vector encoding frequency positional information is added to the input spectrum. The amplitude of the vector is adjusted adaptively and channel dependently\footnote{The amplitude of the vector is adjusted for each channel.} using self-attention. 
This section explores how the adaptiveness and channel dependently of the amplitude coefficient  $\alpha$ affect FAC performance.

The experimental setup is the same as before, with $n_{\text{FAC}}=7$ for FAC-CRNN. Three amplitude adjustment strategies for the encoding vector are considered:
\begin{enumerate}
    \item \textit{fixed}: The amplitude coefficient $\alpha$ is fixed at 1.
    \item \textit{adapt}:  Different channels share the same amplitude coefficients $\alpha$ , which is estimated from the input spectrum.
    \item \textit{adapt}\&\textit{dep}: The amplitude coefficient $\alpha$ is adaptive and dependently adjusted for each channel.
\end{enumerate}
The performance differences between \textit{fixed} and \textit{adapt} highlight the importance of adaptiveness in the amplitude coefficient, while differences between \textit{adapt} and \textit{adapt}\&\textit{indep} demonstrate the impact of channel independence. FAC-CRNN is trained five times under each strategy, with average performance reported.


\begin{figure}[htbp]
	 \centering
	 \includegraphics[width=0.8\linewidth]{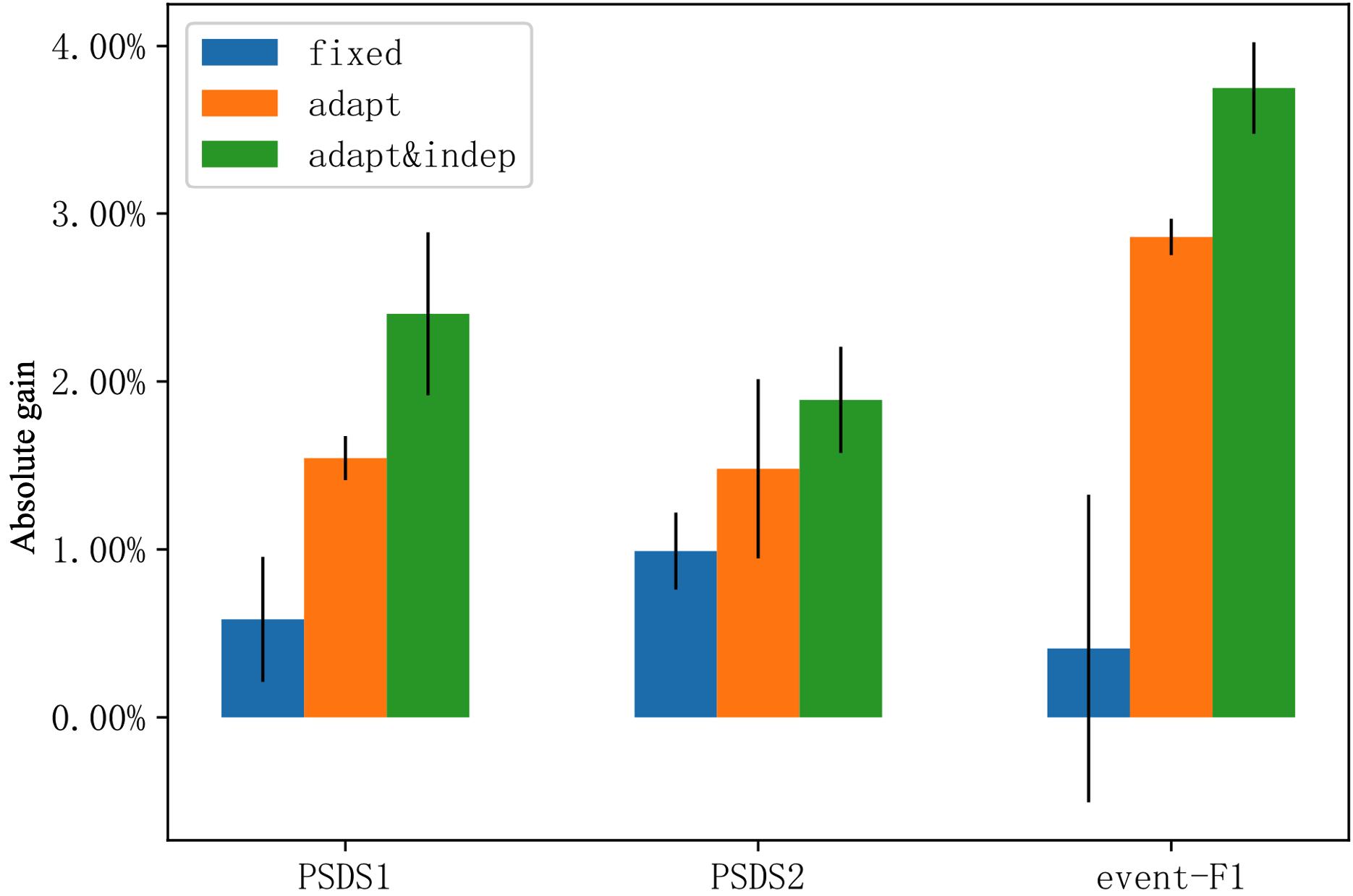}
	\caption{Performance improvements of FAC-CRNN compared to CRNN under three conditions, \textit{fixed}, \textit{adapt} and \textit{adapt}\&\textit{indep}.}
	\label{fig:FAC_adapt-independent}
\end{figure}

Fig. \ref{fig:FAC_adapt-independent} presents the absolute performance improvements of FAC-CRNN over CRNN across the three strategies. \textit{adapt}\&\textit{dep} achieves the highest performance gain, followed by \textit{adapt} with a lesser performance gain, and \textit{fixed} performs the lowest. 
Under the \textit{fixed}, FAC directly adds the positional encoding vector to the input, resulting in a slight PSDS2 improvement over CRNN (average absolute gain of 0.99\%) but similar performance in other metrics. 
After adaptively adjusting the amplitude coefficient (\textit{adapt}), the performance of FAC-CRNN is significantly improved. This result shows that adaptively adjusting the amplitude of the vector is crucial to the performance of FAC-CRNN.
Further introducing channel dependence (\textit{adapt}\&\textit{dep}),  the performance of FAC-CRNN continues to improve slightly, indicating that channel dependence is also important for FAC-CRNN.

\section{Conclusion}

This paper introduces Frequency-Aware Convolution (FAC) for SED. In SED models, CNN layers combined with pooling layers are commonly employed to extract time-frequency patterns. 
However, these CNN-based features may exhibit insensitivity to shifts in the time-frequency patterns along the frequency axis.
To address this issue, FAC encodes the frequency positional information using a vector and adds it into the CNN input. To match the amplitude of input, the encoding vector is adaptively and channel dependently scaled. In contrast, an alternative approach called FDY has also been proposed for the same purpose. Experimental results demonstrate that FAC achieves comparable performance to FDY while requiring significantly fewer parameters and computation. Moreover, ablation experiments prove the criticality of adaptive and channel-dependent scaling in FAC.

 \bibliographystyle{splncs04}
 \bibliography{refs}
\end{document}